\documentclass{aa}  

\usepackage[T1]{fontenc}
\usepackage{graphicx}
\usepackage{natbib}
\usepackage{lscape}

\newcommand{\degree}{\ensuremath{^\circ}}

\def\m2s2{\hbox{\,m$^{2}$\,s$^{-2}$}} 

\def \1s{$1\,\sigma$}

\def \t0{T$_0$}

\newcommand{\prob}[2]{p(\mathrm{#1}|\mathrm{#2})}
\newcommand{\prior}[2]{prior(\mathrm{#1}|\mathrm{#2})}

\newcommand{\project}[1]{\textsf{#1}}

\newcommand{\celerite}{\project{celerite}}

\newcommand{\emcee}{\project{emcee}}

\usepackage{txfonts}
\begin{document} 

   \title{ Improving transit characterisation with Gaussian process modelling of stellar variability}
   \author{  S.~C.~C.~Barros\inst{\ref{IA}}\thanks{E-mail: susana.barros@astro.up.pt}     
          \and O.~Demangeon\inst{\ref{IA}}
          \and R.~F.~D\'iaz\inst{\ref{argentina}}\fnmsep\inst{\ref{argentina2}}
          \and J. Cabrera\inst{\ref{DLR}}
          \and N.~C.~Santos\inst{\ref{IA}}\fnmsep\inst{\ref{UPorto}}
          \and J.~P.~Faria\inst{\ref{IA}}\fnmsep\inst{\ref{UPorto}}
          \and F. Pereira\inst{\ref{IA}}\fnmsep\inst{\ref{UPorto}}
          }
          \institute{ Instituto de Astrof\'isica e Ci\^encias do Espa\c{c}o, Universidade do Porto, CAUP, Rua das Estrelas, PT4150-762 Porto, Portugal \label{IA} 
        \email{susana.barros@astro.up.pt}
              \and
             Universidad de Buenos Aires, Facultad de Ciencias Exactas y Naturales. Buenos Aires, Argentina \label{argentina} 
             \and
             CONICET - Universidad de Buenos Aires. Instituto de Astronom\'ia y F\'isica del Espacio (IAFE). Buenos Aires, Argentina \label{argentina2} 
             \and 
             Deutsches Zentrum fur Luft- und Raumfahrt, Rutherfordstr. 2, 12489 Berlin, Germany \label{DLR}   
                   \and
            Departamento\,de\,Fisica\,e\,Astronomia,\,Faculdade\,de\,Ciencias,\,Universidade\,do\,Porto,\,Rua\,Campo\,Alegre,\,4169-007\,Porto,\,Portugal \label{UPorto}
           }

   \date{Received ??, ??; accepted ??} 
   \abstract
   { New photometric space missions to detect and characterise transiting exoplanets are focusing on
   bright stars to obtain high cadence, high signal-to-noise light curves. Since these missions will be sensitive to stellar
   oscillations and granulation even for dwarf stars, they will be limited by stellar variability. Therefore, it is crucial and timely to develop
   robust methods to account for and correct for stellar variability.}
   { We tested the performance of Gaussian process (GP) regression on the characterisation of transiting
   planets, and in particular to determine how many components of variability are necessary to describe high
   cadence, high signal-to-noise light curves expected from CHEOPS and PLATO. To achieve this, we
   selected a sample of bright stars observed in the asteroseismology field of CoRoT at high cadence
   (32 sec) and high signal-to-noise ratio.}    
   {We used GPs to model stellar variability including different combinations of
   stellar oscillations, granulation, and rotational modulation models. We preformed model comparison
 to find the best activity model fit to our data. We compared the best
   multi-component model with the usual one-component model used for transit retrieval and with a non-GP model.}
   {We found that the best GP stellar variability model contains four to five variability components: one stellar oscillation component, two to four
   granulation components, and/or one rotational modulation
   component, which is consistent with results from asteroseismology. However, this high number of components is in contrast with the one-component GP model (1GP) commonly
   used in the literature for transit characterisation. Therefore, we compared the performance of
   the best multi-component GP model with the 1GP model in the derivation of transit parameters
   of simulated transits.  We found that for Jupiter- and Neptune-size planets the best multi-component GP model is slightly better than the 1GP model, and much better than the non-GP model that gives biased results. For Earth-size planets, 
   the 1GP model fails to retrieve the transit because it is a poor description of stellar activity. The non-GP model gives some biased results and the  best multi-component GP is capable of retrieving the correct transit model parameters. }
   { We conclude that when characterising transiting exoplanets with high signal-to-noise
   ratios and high cadence light curves, we need models that couple the description of stellar variability with the transits analysis, like GPs.
   Moreover, for Earth-like exoplanets a better description of stellar variability (achieved using multi-component models)   improves the planetary characterisation. Our results
   are particularly important for the analysis of TESS, CHEOPS, and PLATO light curves. } 
   

 \maketitle
%

\section{Introduction} 
\label{intro}

Observations of exoplanets are affected by stellar variability. In general the impact is higher for radial velocity observations;  as transit signals are localised in time, they can often
be separated relatively easily from the stellar variability (e.g. CoRoT-7b \citealt{Leger2009,Queloz2009,
Haywood2014, Barros2014} ). However, as the precision of transit observations increases, stellar
intrinsic variability will become the dominant limitation in transit observations of exoplanets,
especially for small planets whose transit depths can be of the order of the amplitudes of stellar
variability (an Earth-size planet orbiting a Sun-like star has a transit depth of $\sim80$ppm). 
The shorter ingress and egress time for Earth-size planets implies that the shape will be more affected by short timescale stellar variability.

Stellar variability has different origins and covers a wide range of timescales. Stars with
convective envelopes show p-mode oscillations with periods of a few minutes and amplitudes $\sim
10$ppm in solar-type stars \citep{Kjeldsen1995,Kallinger2014}. These stars also show photometric
variability due to granulation and super-granulation. Granulation has timescales from $\sim20$
minutes up to days and amplitudes reaching a few hundred ppms \citep{Kallinger2014,Meunier2015}. The
effect of this low amplitude short timescale variability in transit observations has been poorly
studied until now. In contrast, one variability effect that has been well studied is the rotational
modulation due to magnetic activity features on the stellar surface, like  spots, flares, and plages.
These give rise to much higher photometric variations reaching $\sim 1000\,$ppm on a timescale
corresponding to the rotational period of the star \citep{Lagrange2010}. Although the amplitude of
these variations can be high in active stars, their timescale is much longer than the transit
timescale, and hence their effect on the transit shape can be corrected by detrending locally with a
first- or second-order polynomial. However, a residual ambiguity in the determination of the
absolute out-of-flux level can lead to differences in the determination of the planetary radius
\citep{Czesla2009} as seen, for example, in the case of WASP-10b \citep{Christian2009, Maciejewski2011,
Barros2013}. Without resolving the stellar surface the only way to mitigate this effect is to have
 very long baselines of observations. On longer timescales (years), stars vary due to the stellar
magnetic cycle that leads to long-term evolution of spot coverage and produces flux variations up to
a few percent \citep{Baliunas1995,Hall2007,Lovis2011}.

In this work we address how to recover accurate and precise planetary parameters for transiting planets in presence of stellar variability levels representative of current and future space-borne missions. We investigate the effect of stellar oscillations and granulation in transit
observations of exoplanets using the framework of Gaussian processes (GPs). We test several models
that include different components of stellar variability including stellar oscillations, granulation,
and rotation. The advantages of using GPs to model stellar rotational modulation in radial velocity
observations of exoplanets has been shown in many works. For example, \citet{Haywood2014} and \citet{Faria2016} showed
that it is possible to correct the stellar rotational modulation, and to detect planetary signals that
are much smaller than the stellar activity signal. It was also shown that modelling stellar activity with GPs
in photometric observations of exoplanets allows us to correct for these factors \citep[e.g.][]{Aigrain2015,
Serrano2018}. However, the stellar rotational modulation signal is quasi-periodic and since we can sample the typical timescale
with enough data it is possible to make good predictions. In contrast, granulation is a
stochastic process and making predictions is much harder. The stellar oscillations are also quasi-periodic,
but contain many modes which  might lead to worse predictions. Hence, it is not clear whether a combined
model of the different variability components will improve transit parameter determination. We expect that the combined model will account for the uncertainty introduced by stellar variability increasing the accuracy of the parameters even at the expense of precision. Perhaps more interesting is whether it can correct stellar variability and increase the precision of the parameters. Another question we address is how many components of stellar variability are
necessary to model the light curves.

Photometric variability due to stellar oscillations and granulation was detected in Kepler
observations \citep{Borucki2010,Mathur2011, Bastien2014,
Cranmer2014,Kallinger2014}. However, these were relatively rare due to the long cadence of Kepler light curves
and average magnitude of the Kepler field stars. Hence, to study how stellar oscillations and
granulation affect transit observations, we chose bright stars observed in the asteroseismology
field of CoRoT \citep{Baglin2006}. We want to  test whether GPs can correct or account for stellar
variability in high cadence high signal-to-noise light curves, and to determine  the best model to use. We
start by presenting our sample in Section~2, followed by the presentation of our variability model
and model comparison methods in Section~3. In Section~4 we show how we derived the best stellar
variability GP model. In Section~5 we determine in which cases the multi-component model is required.
Finally, we discuss the implication of our results in Section~6.

%
%
\section{Stellar sample}

\begin{table*} 
\centering 
\caption{Characteristics of the sample of stars: spectral type, V magnitude, rotation period ($P_{rot}$ ), and presence of a known planet
\label{table:stars}}
\centering \begin{tabular}{l  l   c c  l} 
\hline 
\hline 
Star &  Spectral type &  V mag  & $P_{rot}$ & Known planet \\ 
\hline HD 43587   &  G0 V  &  5.7 & long & \\ 
HD 49933   &  F3 V   & 5.8 & 3.4 & \\ 
HD 52265   &  G0 V  &  6.3 & 12.3 & non-transiting Jupiter\\ 
HD 179079 &  G5 IV &  8.0 & &non-transiting warm Neptune \\ 
\hline 
\hline
\end{tabular} 
\end{table*}

\label{Observations} 

\subsection{CoRoT light curves} 

The CoRoT satellite had two science channels
for its two science goals: asteroseismology and exoplanet search. Each initially had two CCDs, which
were reduced to one CCD per field after the failure of the Data Processing Unit 1 on 8 March 2009. 
Due to pointing restrictions CoRoT observations were divided into long runs with a duration up to 150
days and short runs with a duration of $\sim$30 days.

In the exoplanet field, to increase the probability of detecting transiting exoplanets 
$6\,000-12\,000$ target stars (magV > 9) were monitored in each run, the majority with a cadence of
512 seconds. In the asteroseismology channel to reach the signal-to-noise ratio necessary to detect stellar oscillations,
only a few very bright stars were observed (average V magnitude of 7) at a much higher sampling rate (1 second).
 The asteroseismology channel had five stellar windows (50x50 pixels), five sky-reference windows, and two offset
reference windows per CCD. Aperture photometry was performed on board using a mask optimised for the
position in the CCD. Several corrections to the data were applied to correct for
instrumental effects, and the data was resampled to 32 seconds in the heliocentric frame. In particular, several steps of outlier rejection were preformed to clean cosmic ray hits, first in the images and then in the light curves. We used the
latest reduction of the light curves that is available through the CoRoT legacy archive
\footnotemark \footnotetext{$https://corot.cnes.fr/en/release-corot-legacy-data$}. No extra outlier rejection was preformed.  A full
description of the CoRoT satellite can be found in \citet{Auvergne2009}, while a more recent review can be found in \citet{2016corotbook}.

Since we are interested in observations at high cadence and with a very high signal-to-noise ratio, we
selected observations made by the asteroseismology field of CoRoT during long runs.  Our sample contains three main sequence stars with magnitudes between
5.7 and 6.3 and photometric precision between 56 and 84 ppm over 32 seconds bin, and a slightly fainter sub-giant star
 (mag=8.0) for comparison. This sample allowed us to probe the timescales and
amplitudes that we are interested in.  We included targets with known non-transiting planets so that our sample is
representative of planet hosts. The properties of our four stars are given in Table~\ref{table:stars},
while the details of the observations are given in Table~\ref{table:observations}. In this table we also give the combined differential photometric precision (CDPP)-6.5 hours calculated following the method of \citet{Gilliland2011} for comparison with the Kepler sample. The CDPP-6.5 only measures the intrinsic variability of stars on timescales between 6.5 hours and 2 days. For all the dwarfs in our sample we obtained a CDPP-6.5 hours lower than 5 ppm. Hence, our stars have low intrinsic variability when compared with Kepler dwarfs stars \citep{Gilliland2011} and also when compared to the Sun.  Therefore, these low variability stars are representative of the best targets for transit search with future missions. For the sub-giant star HD~179079 the CDPP-6.5 hours is higher, as expected, (11.2 ppm) and will be used as comparison.

\begin{table} 
\centering 
\caption{Details of the observations of each star including the name of the CoRoT run, the total duration of the observations, and the uncertainties per 32-second
bin.
\label{table:observations}} 
\centering 
\begin{tabular}{l  l  c c  c } 
\hline 
\hline Star &  run & duration  & $\sigma$ (/32s) & CDPP 6.5 h \\ 
& &  (days)  & (ppm)  & (ppm) \\ 
\hline 
HD 43587   & LRa03   & 142  & 56  & 4.3\\ 
HD 49933   & IRa01    & 61 & 57 &  4.4\\ 
HD 49933   & LRa01    & 137 &  56 & 4.1\\ 
HD 52265   & LRa02  & 117  &  84  &4.6 \\ 
HD 179079 & LRc09  & 55 & 170  & 11.2 \\ 
\hline 
\hline 
\end{tabular} 
\end{table}

 \subsection{Previous asteroseismology analysis}

As main targets of the CoRoT asteroseismology program, results of the asteroseismic analysis of most of these stars were already reported in the literature.
 The light curve of HD~43587 was analysed by  \citet{Boumier2014} who measured  p-mode oscillations
with frequency peaking at $2247 \pm 15\, \mu Hz$ (7.42 minutes).
The light curve of HD~49933 was analysed by \citet{Appourchaux2008} and \citet{Benomar2009} who
measured solar-type oscillations with a central frequency of $1760\, \mu Hz$ (9.50 minutes) and a
rotation period of 3.4 days.   
 The light curve of HD~52265 was analysed by \citet{Ballot2011} who reported solar-type
 oscillations in the range $1500 - 2550\, \mu Hz$  with a central frequency of $2090 \pm 20\, \mu
 Hz$ (7.97 minutes). They also reported one granulation component with a period of $4.03\pm 0.03$
 minutes and rotational modulation with a period of $12.3 \pm 0.15$ days with signs of differential
 rotation. No analysis was published from HD~179079 CoRoT observations.

\section{Variability modelling and model comparison}

\subsection{Gaussian process regression}

Gaussian processes (GPs) are non-parametric models that are useful for cases where the functional form of the
model is not known a priori \citep{Rasmussen2006}. GP models have been used for Bayesian regression to model instrumental systematic noise \citep{Gibson2012} and stochastic processes.
Recently they have been also used to model stellar activity \citep{Haywood2014,Aigrain2015}. The
form of a GP is defined by a mean function and a covariance matrix, which is modelled by a kernel function. There are
several classes of GP kernels which describe different behaviour for the correlation between data
points. An advantage of the Bayesian framework is that it penalises complex
models and hence avoids overfitting.

One disadvantage of GPs is that  the computation time generally scales with the number of
observations cubed ($\mathcal{O} (n^3) $). Fortunately, a new implementation of GPs has recently been developed
 called \celerite\  \citep{Foreman-Mackey2017}, which considerably speeds up computation time
as it scales with $\mathcal{O}  n$. This implementation comes with some restrictions; for example,
it can only be applied to one-dimensional datasets  and requires stationary processes. This means
that the kernels are required to be functions of $\tau$ alone, with $\tau_{ij} = |t_i - t_j|$. It
also requires the kernels to be a mixture of exponential functions. \citet{Foreman-Mackey2017} show
that they can be re-written as a mixture of quasi-periodic oscillators. Furthermore, what is most
interesting for our case is that some of the possible \celerite\ kernels are well suited to describe
different forms of stellar variability. According to equation 49 in \citet{Foreman-Mackey2017}, the
kernel for a stochastically driven, damped harmonic oscillator with a quality factor (Q ) larger
than 0.5 is given by 

\begin{equation} 
\label{sho-kernel} 
k(\tau;\,S_0,\,Q,\,\omega_0) = S_0\,\omega_0\,Q\,e^{-\frac{\omega_0\,\tau}{2Q}}\, \cos{(\eta\,\omega_0\,\tau)} + \frac{1}{2\,\eta\,Q} \sin{(\eta\,\omega_0\,\tau)} 
,\end{equation} 
where $\omega_0$ is the frequency
of the undamped oscillator,  $S_0$ is related to the power spectral density (PSD) at $\omega = \omega_0$ by
$S_0  = PSD(\omega_0) / ( \sqrt{2/\pi}\,Q^2 ),$ and $\eta = \vert 1-(4\,Q^2)^{-1}\vert^{1/2}$.

For the particular case of $Q = 1/\sqrt{2}$ this kernel has the same power spectrum density as
stellar granulation \citep{Harvey1985, Kallinger2014} and can be rewritten as 
\begin{equation}
k(\tau) = S_0\,\omega_0\,e^{-\frac{1}{\sqrt{2}}\,\omega_0\,\tau}\, \cos{\left(\frac{\omega_0\,\tau}{\sqrt{2}}-\frac{\pi}{4}\right)} \quad 
\end{equation} 
(equation 51 in \citealt{Foreman-Mackey2017}). Therefore, we use this kernel to describe stellar granulation and
we refer to it as the granulation kernel. In the classical GP framework this is close to the
square exponential kernel which was previously used to model stellar activity together with transit modelling
\citep[e.g.][]{Dawson2014, Barclay2015}.

For the limit of Q > 1, the kernel given by equation 1 has a Lorentzian power spectrum density near
the peak frequency. Therefore, it can be used to describe stellar oscillations. We also use this
model and we refer to it as the oscillation kernel.

To model stellar variability due to rotation modulation of spots and plages it is common to use the
quasi-periodic kernel  in radial velocity modelling \citep{Haywood2014} and in photometry
\citep{Aigrain2015,Serrano2018}. A kernel with a similar covariance function in \celerite\ was
proposed by \citet{Foreman-Mackey2017} (Eq.  61), 
\begin{eqnarray} 
\label{rot-kernel} 
k(\tau) = \frac{B}{2+C}\,e^{-\tau/L}\,\left[ \cos\left(\frac{2\,\pi\,\tau}{P_\mathrm{rot}}\right) + (1 + C) \right] 
,\end{eqnarray} 
where $P_\mathrm{rot}$ is the rotation period of the star and $B>0$, $C>0$,
and $L>0$. We use this kernel to model stellar rotation modulation in the light curves, which we refer to as
the rotation kernel.

The objective of this study is to test which of these variability components can be detected in our
light curves, and to determine their significance. To test this, we construct several noise models that include
different components of granulation, oscillation, and rotation by adding the respective covariance
matrixes (kernels) described above. We call them noise models because they model the covariance and
not directly the data. In our case the deterministic model is zero in the first part of this work (section 4), while for the second part it is the transit model (section 5). The parameters of the GP are
called hyper-parameters to distinguish them from the transit model parameters. The different
combinations of the noise models considered will be presented in section~4. For each noise model, we
find the best hyper-parameters for each light curve maximising the log-likelihood function

\begin{eqnarray} 
\mathrm{ln}\, \mathcal{L} ( \bold{r})= -\frac{1}{2}\bold{r}^{\mathrm{T}} (\bold{K}+\sigma_i^2\bold{I})^{-1} \bold{r}  -\frac{1}{2} \mathrm{ln}(|\bold{K}+\sigma_i^2\bold{I}|) -\frac{n}{2} \mathrm{ln}  (2\pi) 
,\end{eqnarray}
where r are the residuals obtained by subtracting the deterministic model to the data, $|\bold{A}|$  is the determinant of the matrix A, and n is the
number of data points. The term $\sigma^2\bold{I}$ represents an additional white noise component,
where $\bold{I}$ is the identity matrix and $\sigma^2$  is the variance of the extra noise.

We find the maximum of the log-likelihood function using the Markov chain Monte Carlo (MCMC) algorithm \emcee\
\citep{Foreman-Mackey2013}. We used 32 chains, which is  double  the maximum number of
parameters fitted (16). Each chain has 10000 iterations.  We separated the exploration in
three stages. The first stage was used to explore the parameter space and find the global maximum and consisted of 2000 steps.
The chains were started at random points from the prior.  The second
stage was used to consolidate the global maximum and consisted of 4000 steps. We started the chains
close to the parameter set with highest posterior probability found in the first stage. The third stage
was used to explore the parameter space next to the global maximum and derive the best
hyper-parameters and their uncertainties. It was just a continuation of the second stage, but only
this last stage was kept and analysed. Convergence was checked with the Geweke algorithm
\citep{Geweke1992} and when necessary some residual burn-in was discarded from the third stage. The chains were
combined in a master chain that was used for further analysis. To infer the parameter values we used
the median of the master chain distributions for all the hyper-parameters except for the stellar
rotation period.  The median is usually a better estimator than the mode, which is very sensitive to
bin size. However, for distributions with large tails and high asymmetry (in our case the stellar rotation period)
the median is a poor estimator and the mode is better. Hence, in these particular cases we used the
mode of the distribution. In some cases the stellar rotation period could only be constrained to be
above a certain value, and in these cases we quote the $3\sigma$ limit. The uncertainties of the
hyper-parameters were estimated from the $16{th}$ and $84{th}$ percentiles of the chains.

For the GP noise models, we used wide priors for the hyper-parameters. For comparison with asteroseismology, we converted
the hyper-parameters of equations 1 and 2 into the parameters usually used in asteroseismology
analysis to fit the power spectrum following \citet{Pereira2019}:

\begin{align}
a_{gran} =\sqrt{  \sqrt{2} S_0 \omega_0 }\\ 
a_{osc}  = 4 S_0 Q^2 \\ 
\tau= \frac{ 2 \pi}{\omega_0}
\end{align}

The priors are the same for each component of variability (i.e. no order was imposed on the granulation timescales). This was meant to
simplify the exploration of the parameter space. In cases where the second component of granulation
had the same period as the first within the errors, we considered that the second component  was not
needed. The priors used for the oscillation kernel and the rotation kernel are given in
Tables~\ref{prior_granulation} and \ref{prior_rotation}. The granulation kernel is a particular case of
the oscillation kernel where  $Q = 1/\sqrt{2}$ and the rest of the hyper-parameters have the same
priors as the oscillation kernel.

\begin{table} 
\caption{Priors for the granulation and oscillation kernels \label{prior_granulation} }
\begin{tabular}{lcc} 
\hline 
\hline Parameter & Prior \\ 
\hline ln white noise  &  $\mathcal{U}(-15; 5)$ \\ 
$ln S_0$    & $\mathcal{U}( -5; 25)$ \\ 
$ln \omega_0$ & $\mathcal{U}(-1; 8)$ \\ 
$ln Q$  & $\mathcal{U}(-5; 2.35)$ \\ 
\hline \hline \end{tabular} 
\\ $\mathcal{U}(a;b)$ is a uniform probability distribution between $a$ and $b$. 
\end{table}

\begin{table} 
\caption{Priors for the rotation kernel \label{prior_rotation} }
\begin{tabular}{lcc} 
\hline 
\hline Parameter & Prior \\ 
\hline ln white noise  &  $\mathcal{U}(-15; 5)$ \\ 
ln a  & $\mathcal{U}( 7; 12)$ \\ 
ln b & $\mathcal{U}(-1; 0)$ \\ ln c & $\mathcal{U}(-4.5; 0.5)$ \\ 
ln period & $\mathcal{U}(0.5; 5)$ \\ 
\hline 
\hline 
\end{tabular} \\ 
$\mathcal{U}(a;b)$ is a uniform probability distribution between $a$ and $b$. 
\end{table}

\subsection{Model comparison} 

\label{modelcomparison} 
Bayesian probability theory allows us to
perform model comparison by the computation of the odds ratio between two hypotheses
\citep[e.g.][]{Diaz2014}. The odds ratio for a pair of hypotheses is the multiplication between the prior odds
and the Bayes factor. The prior odds are the {a priori} probability of each model. In our
study we assume that the prior odds are equal for all the models as different stars are dominated by
different types of variability and the noise of the data will affect the detection of the
variability components in a way not known {a priori}.  
Therefore, in our case, the odds ratio is equal to the Bayes factor
which is the ratio of the two evidence terms. The evidence of a model is given by the integral of the joint
posterior of the model's parameters
\begin{equation} 
\prob{D}{H_i, I} = \int{  \prob{D}{\bold{\theta_i}, H_i, I}\cdot\mathrm{d}\bold{\theta_i} \cdot   \prior{\bold{\theta_i}}{H_i, I}   }\;\;,
\label{eq.evidence} 
\end{equation}
where $D$ represents the data, $H_i$ the hypothesis $i$, $I$  the prior information,
$\bold{\theta_i}$  the parameter vector of the model associated with hypothesis $H_i$,
$\prior{\bold{\theta_i}}{H_i, I}$  the joint prior distribution, and
$\prob{D}{\bold{\theta_i}, H_i, I}$  the likelihood for a given dataset $D$ under the assumption
of hypothesis $H_i$. Bayesian model comparison penalises models with a larger number of parameters
because they dilute the normalised prior density, leading to a natural occam's razor.

The priors are given in Tables~\ref{prior_granulation} and \ref{prior_rotation}, and the likelihood is
computed with equation 4. This multi-dimensional integral is in general impossible to
compute analytically, and several approximations are used. In our case we used importance sampling
\citep{Kass1995} to approximate the integral, and in particular the Perrakis method
\citep{perrakis2014}.  This method uses samples of marginal posterior probability distributions from an MCMC
algorithm to estimate the evidence.

To estimate the evidence we used 3000 independent samples of the posterior probability
distribution. For each model this procedure was repeated 400 times in order to obtain the
distribution of the estimator of the evidence. The value of the evidence we quoted is the median of
this distribution. More details about our computation of the evidence can be found in
\citet{Diaz2014} and \citet{Nelson2018}.

\subsection{Stellar rotation}

From the light curves (Figures~\ref{lcmodel}, \ref{lcmodelzoom} ) it is evident that HD~49933
and HD~52265 show a clear sign of rotation spot modulation. This implies that some light curves
require a model that includes a rotation kernel. Preliminary tests showed that it was difficult to
fit the stellar rotation using our rotation \celerite\ kernel (equation 3) due to the strong short-term variability present in the light curves that is also reproduced by this Kernel. This was  solved once more kernels were added
to the GP model to account for short-term variability.  However, we found that for clarity it was easier to start by dividing our sample into
stars with and without a measurable rotation period. 

To determine for which light curves it was
possible to measure the rotation period we averaged out high frequency variations by binning the
light  curves into two-hour bins. The binned light curves were fitted with the method described
above with one rotation kernel. The derived rotation period for each of the observation is given in
Table~\ref{table:rotationp}. We derived the rotation period for HD~49933, HD~52265, and HD~179079.
However, we could only put a lower limit on the rotation period of HD~43587. 

\begin{figure*} 
\centering 
\includegraphics[width=0.90\textwidth]{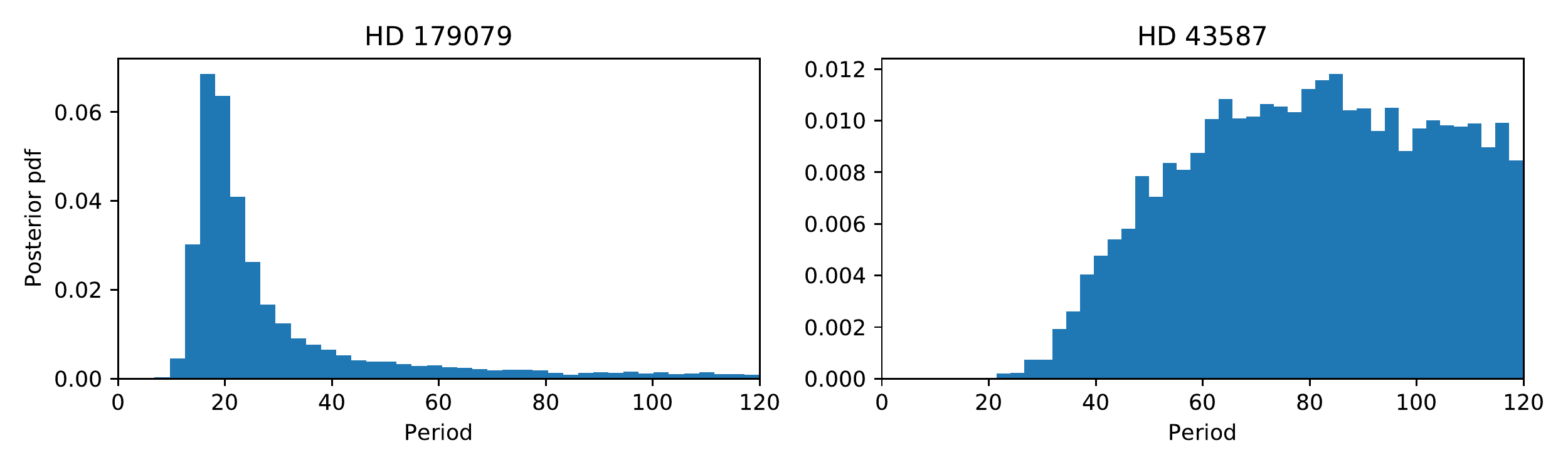}
\caption{Derived posterior probability distributions for the stellar rotation period for two
examples: HD~179079 where the rotation period is well determined, and HD~43587 where the rotation
period is longer than 29.13 days at $3 \sigma$. 
\label{fig.chain}  } 
\end{figure*}

The rotation periods  were already derived from these light curves by \citet{Appourchaux2008} and \citet{Benomar2009} for HD~49933,  and by \citet{Ballot2011} for HD~52265,   as mentioned in Section 2.2. Our values are in agreement with the published values. In
Figure~\ref{fig.chain} we show the posterior probability distribution of the rotation period for two
examples, one where we can determine the rotation period (HD~179079) and one where we can only
derive a lower limit (HD~43587). It should be noted that  for HD~49933 the derived rotation
period for the two observations agrees well. Since we could not determine the rotation period for HD~43587, for this star we did not consider models that include a rotation kernel for the subsequent analysis. Hence, we separated HD~43587 from the main group of stars for which we detect the rotation period.

\begin{table} 
\centering 
\caption{Derived stellar rotation period for the binned light curves. \label{table:rotationp}} 
\centering 
\begin{tabular}{lc} 
\hline 
\hline Star &  rotation Period \\ 
&  (days)   \\ 
\hline
HD 43587   &  $>  29.13$ \\ 
HD 49933 (IRa01)  &  $3.19 \pm 0.16 $ \\ 
HD 49933 (LRa01)  & $3.294\pm 0.082$ \\ 
HD 52265   & $10.3\pm 1.2$  \\ 
HD 179079 & $17.8^{+29}_{-1.6}$\\ 
\hline 
\hline
\end{tabular} 
\end{table}

\section{What is the best stellar variability GP model?} 

All the light curves were fitted with a set
of models combining granulation, oscillation,
and rotation when needed (see section 3.3).
The notation used to designate a model summarises the number of granulation, oscillation, and rotation
kernels used for a given model. We used the letter G for granulation, O for oscillation, and R for
rotation. The number following the letter indicates how many of these kernels are considered. For
example, the G5 model is composed of 5 distinct granulation kernels, the G4O1 model is composed of 4
granulation kernels and 1 oscillation kernels.

As we can build an infinite number of models from our three building blocks, we limited our set of
models according to the following criteria:
\begin{itemize}
\item We limited the total amount of components to five according
to the maximum number of components considered in the literature \citep{Harvey1985, Corsaro2015}. This limits the computational time to a considerable but still manageable amount (33 days per target). 
\item We limited the number of oscillation kernels to one since we expected only
one oscillation component in the light curves.
\item We limited the number of rotation kernels to one since we expected only
one rotation component in the light curves.
\item We did not limit the maximum number of granulation components to three, as expected, because we
assumed this kernel is also capable of detecting other types of variability (instrumental or stellar),
and we considered it to be the simpler model as it has fewer parameters. The
classical squared exponential kernel (which is the one closer to the granulation \celerite\  kernel) is
the  most commonly used in the literature to model both instrumental and stellar red noise in light curves. We started by considering models made only with granulation components (up to
five). We then replaced some of these components by oscillation or rotational components. 
\end{itemize}

The final sets of
models that were considered are given in Tables~\ref{comprotation} and \ref{components} for the
stars with and without detected rotational modulation respectively. Tables~\ref{comprotation} and \ref{components} 
also present the differences between the logarithm of the evidence
(Section~\ref{modelcomparison}) of the best model  and all the other models considered. We analysed the
fits for convergency and made sure that the components were different. When two components 
effectively had the same timescale the fit was not considered for model comparison. This was the case of the model G4O1 for both light curves of HD~49933  and the light curve of HD~179079, where one of the components had the same timescale as another component.

\begin{table*} 
\caption{ Differences in the log evidence of each model relative to the best GP noise model for targets with measured stellar rotation period. \label{comprotation} }
\begin{tabular}{lcccc} 
\hline 
\hline 
Model & HD 49933 IRa01 &  HD 49933 LRa01 & HD 52265 & HD 179079 \\ 
\hline 
G1      &       -6776   &       -18489  &       -11223  &       -1768   \\
G2      &       -786    &       -2270   &       -1514   &       -222    \\
G3      &       -161    &       -478    &       -344    &       -130    \\
G4      &       -125    &       -240    &       -152    &       -17     \\
G5      &       -112    &       -211    &       -144    &       -18     \\
G1R1    &       -251    &       -574    &       -845    &       -125    \\
G2R1    &       -121    &       -188    &       -137    &       -18     \\
G3R1    &       -112    &       -181    &       -136    &       -13     \\
G2R1O1   &      \textbf{0}      &       \textbf{0}      &       \textbf{0}      &       \textbf{0}      \\
G3O1    &       -9      &       -84     &       -56     &       -6      \\
G4O1    &       -       &       -       &       -16     &       -       \\
\hline 
\hline 
\end{tabular} 
\end{table*}

\begin{table} 
\caption{ Difference in the log evidence of each model relative to the best GP noise model the target with unmeasured stellar rotation period. \label{components}}
\begin{tabular}{lc} 
\hline 
\hline 
Model  & HD 43587 \\ 
\hline
G1      & -11256\\ 
G2      & -1847\\ 
G3      & -652\\ 
G4      & -369\\ 
G5      & -345\\ 
G3O1& -66\\ 
G4O1 &\textbf{0}\\ 
\hline 
\hline 
\end{tabular} 
\end{table}

\begin{figure*} 
\centering 
\includegraphics[width=0.99\textwidth]{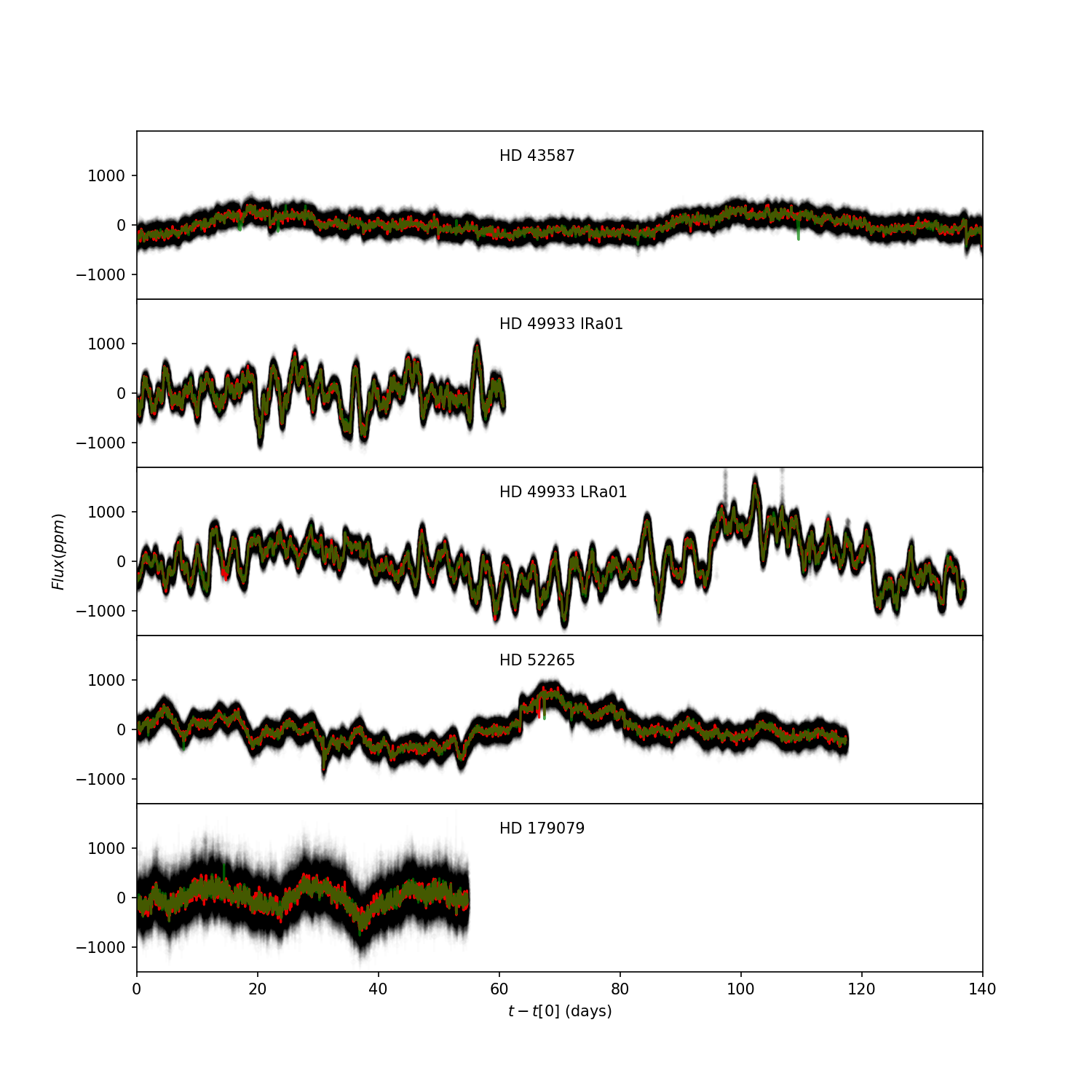} 
\caption{Light curves of all our sample stars overplotted with the best GP model in red and the G1 model
in green. For clarity we decreased the transparency of the light curve points. The x and y axes are
the same for all the observations for easier comparison between the different observations.  \label{lcmodel}} 
\end{figure*}

\begin{figure*} 
\centering 
\includegraphics[width=0.99\textwidth]{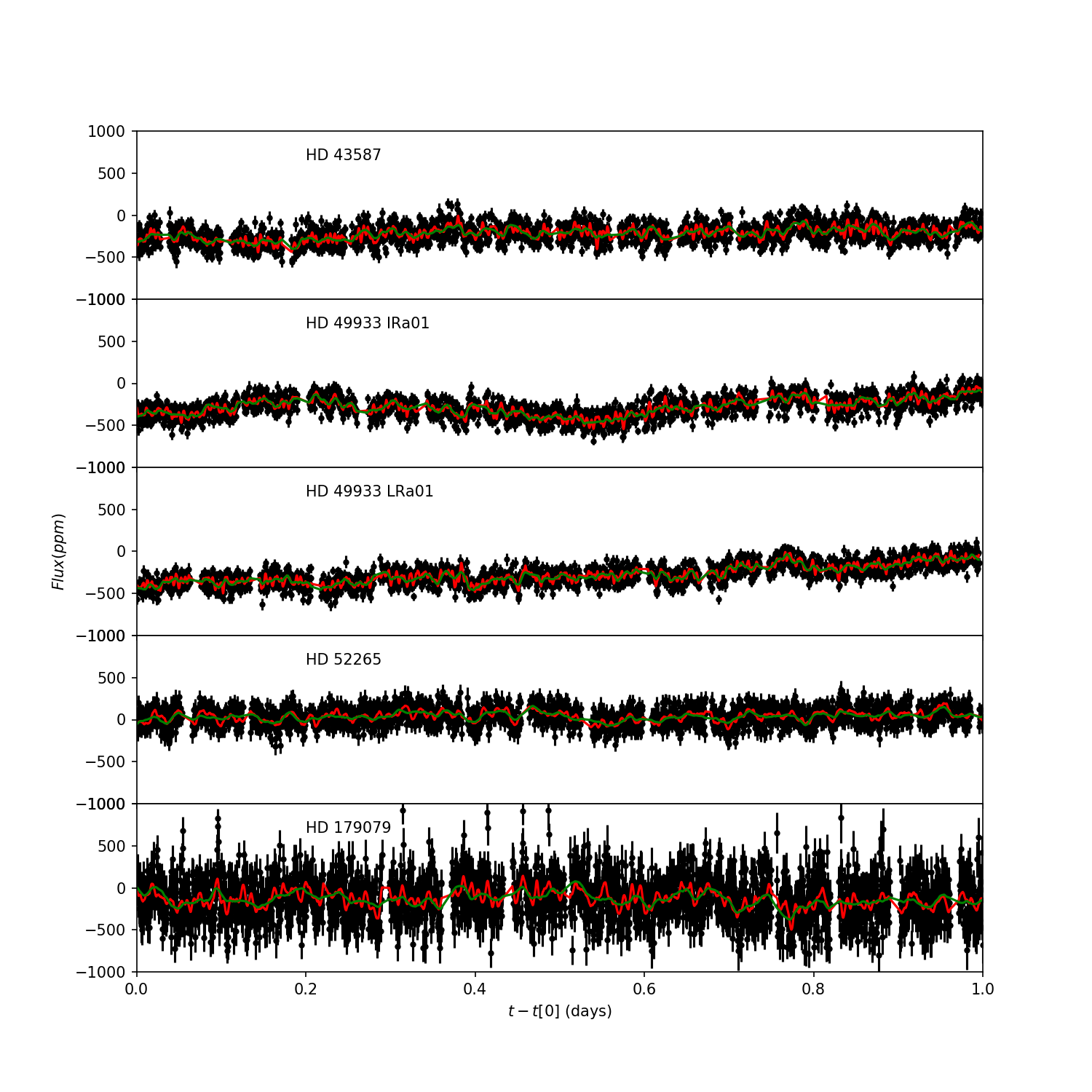}
\caption{ Zoom of the first day of observations of all our sample stars shown in Figure~\ref{lcmodel}.
In the zoom it is clearer that the best model (red) has a much higher frequency component than the
G1 model (green).  \label{lcmodelzoom}} 
\end{figure*}

To identify the best model, we use the classical criteria of  \citet{Jeffreys1961} stating that a more complex model is only considered as better if it is associated with  evidence 150 times
higher than the simpler model, which corresponds to a delta log evidence superior to five. According to
this threshold the best GP noise models are G2R1O1 for all of the light curves with detected rotational modulation (HD~49933, HD~52265,  and HD~179079) and G4O1 for HD~43587. It is worth noting that  the results are the same for the two light curves of HD~49933. Importantly, the difference between the best GP noise model and the GP noise model
with just one kernel (G1), which is usually used in the literature, is highly significant. This implies that a
multi-component model is really needed when modelling the variability of these stars. Given that it is
common practice to use just one kernel to model the photometric stellar variability in transit
parameter retrieval, it is interesting to further compare the best GP noise model we found with the G1
model. In Figure~\ref{lcmodel} we show the full light curves for each star and we overplot the best
model (red) and the G1 model (green). In Figure~\ref{lcmodelzoom}, we show a one-day zoom of the
previous figure so that the difference between models on the shorter timescales is clearer). From
the figures it is evident that the G1 model reproduces well the long-term  variability of the light
curves, but does not reproduce the short-term variability.

 In Table~\ref{table:k1model} we give the fitted hyper-parameters for all the light curves when we
 considered the G1 model, and in Tables~\ref{table:k3rotmodel} and \ref{table:k3model} we give the
 fitted hyper-parameters for the best GP noise model. The timescales found for the G1 model are
 between 20 and 40 minutes.  The timescales found for the best model are close to 8 minutes for the
 oscillations (14 minutes for the sub-giant HD~179079),  7-27 minutes for the first granulation kernel (43 minutes for the sub-giant HD~179079), and up to 14 hours for the longer
 period granulation kernel.  In general,  the amplitudes of variability are higher for the rotational component as
 expected.

\begin{table*} 
\centering 
\caption{Derived hyper-parameters for the G1 model. 
\label{table:k1model}}
\centering 
\begin{tabular}{lccc} 
\hline 
\hline Star &  wn  &amplitude & period\\ 
& ppm &  ppm &min   \\ 
\hline 
HD 43587 & 4.2356$\pm$0.0022    &       209.1$\pm$2.2   &       20.97$\pm$0.30\\ 
HD 49933 (IRa01) &      4.0466$\pm$0.0040       &       438.0$\pm$8.5   &       32.22$\pm$0.65\\ 
HD 49933 (LRa01) &              3.9596$\pm$0.0029       &       627.8$\pm$8.5   &       35.45$\pm$0.46\\ 
HD 52265  &     3.9284$\pm$0.0053       &       384.6$\pm$6.1   &       39.91$\pm$0.74\\ 
HD 179079       &       5.1370$\pm$0.0039       &       258.4$\pm$4.8   &       24.87$\pm$0.76\\ 
\hline 
\hline
\end{tabular} 
\end{table*}

\subsection{Comparison with previous asteroseismology analysis}

In order to validate our method we compared our results with previous results from asteroseismology.
The majority of the previous analysis of these light curves using frequency domain methods only report the timescales of the
stellar oscillations. Stellar oscillations were found for all of the light curves previously analysed. 
In our analysis, models that include one oscillation kernel were strongly preferred for all the stars, and hence we also detect stellar oscillations in all the light curves.

CoRoT light curves have gaps due to the crossing of the South Atlantic anomaly. Since our method
uses GPs that are applied in the time domain, it is not affected by gaps contrary to frequency domain analysis. For asteroseismic analysis, it is usual to fill these gaps
in order to apply the Fourier domain methods. Since even the best
interpolation methods will alter the data, we do not perform gap filling in our analysis. However, this might lead to small differences between our analysis and previous asteroseismology results.

Our derived timescales of the stellar oscillations are similar to the published values even if they are not strictly consistent (taking into account error bars). The small differences between the two analyses might be due to specific data reduction procedures, gap filling methods, or the number of components considered to fit the data.  
A comparison between GPs and asteroseismology methods to analyse stellar granulation and oscillations for red giant stars was preformed by  \citet{Pereira2019}. They show that both models find the same stellar signals, but there are some slight offsets in the derived parameters due to a difference in the shape of the models used in asteroseismology and the GP kernel used.

For the light curves with detected rotational modulation, GP models that include the rotational kernel are preferred.

Our derived rotation periods are in good agreement with previous reported values derived for HD~49933 and HD~52265. For HD~43587, although we do not detect a rotation period, the timescale of the longer granulation component ($15.4\,$days) is much longer than expected for stellar granulation. Hence,  this timescale is probably related with stellar activity, either spot rotational modulation or the timescale of emergence of active regions.

\section{Transit parameter retrieval}

Very few transit analysis studies have included GPs to model stellar variability. The previous GP
implementations were so slow that applications to large datasets were almost prohibitive. The
exceptions are mostly evolved stars because of their longer variability timescales and higher variability amplitude,  for example Kepler-91\citep{Barclay2015},  Kepler-419 \citep{Dawson2014}, and K2-97b \citep{Grunblatt2016}. These stars were successfully analysed using GPs, but only one GP component was considered. In previous observations with the CoRoT
and Kepler satellites, multiple-component noise models were not required due to a combination of the
low signal-to-noise ratio and lower cadence. For example, for the long cadence of Kepler the shorter timescales would be averaged out leaving just the rotational modulation, and possibly the long-period
granulation in the light curve. Many methods were used to detrend the rotation modulation in light
curves including, albeit rarely, GPs. Moreover, most of the short cadence light curves did not have a high
enough signal-to-noise ratio to detect the small amplitudes of the oscillations and granulation ($ < 50$
ppm Tables~\ref{components} and \ref{comprotation}).

 As we show in the previous section, for high
signal-to-noise observations taken at high cadence (30 sec) a larger number of components is needed
to correctly account for the variability in the light curves. However, when we are interested in transit
parameter retrieval it is more important  to determine whether the higher number of components would
lead to a more accurate derivation of transit parameters. To test this, we injected planetary
transits in the above light curves and derived the transit parameters comparing the one-component
 model with the best model found in the previous section. For completeness we also compare them with transit parameters derived with a classical non-GP model. For simplicity, for HD~49933 for which we have two light curves, we considered only the LRa01 light curve because it is longer than IRa01.

\subsection{Transit injection}

In this work we tested a general case of space-based observations coming from transit surveys like
CoRoT, Kepler, TESS, or PLATO \citep{Baglin2006, Borucki2010, Ricker2015, Rauer2014} where several
consecutive transits are available.  In order to have several transits in our light curves, we
chose to inject planets with an intermediate period of 15 days.  It will also be interesting to test
transits with very short periods and very long periods to test the impact of different regimes of
stellar activity in the long term (e.g. rotational and magnetic cycles). Furthermore, it would also
be interesting to test observations where just one transit is available, which will be especially relevant for
the search of Earth-like planets with TESS or PLATO. However, we leave these analyses of specific cases to future work.

To test different transit signal-to-noise regimes, we injected transits of a Jupiter-size planet
(same mass and radius as Jupiter), a Neptune-size planet, and an Earth-size planet. We assumed a Sun-like star and  the same quadratic limb darkening coefficients LD1 = 0.5048 and LD2 = 0.1468 for all of
the light curves (corresponding to WASP-18 as a random example).  We also assumed central transits (
 $inc$ = 90 \degree ) and circular orbits. For each planet, the simulated normalised separation of
the planet ($a/R_{\star}$) and the planet-to-star radius ratio ($r_{p}/R_{\star}$) take into
account the mass and radius of the planet and the star and the orbital period. The mid-transit time was set to be 5 or 6 days after the beginning of the observation of each light curve t[0].
The full set of simulated parameters is provided in Table~\ref{injected}. We used the package \project{batman}  \citep{batman2015} to
simulate and model the transits.

\begin{table} 
\caption{Simulated transit parameters for each type of planet. We also assumed an orbital period of 15 days,  $inc$ = 90 \degree, circular orbits, and quadratic limb darkening coefficients LD1 = 0.5048 and LD2 = 0.1468 .\label{injected} }
\begin{tabular}{lccc} 
\hline 
\hline 
Planet   & T$_{0} $[d]  &$r_{p}/R_{\star}$ & $a/R_{\star}$  \\
\hline 
Jupiter  & t[0] + 6 &  0.07419688 & 18.960539 \\ 
Neptune  & t[0] + 5 & 0.02581571 & 18.955250 \\ 
Earth  & t[0] +  6&     0.00661941  & 18.954969 \\ 
\hline 
\hline 
\end{tabular} 
\end{table}

\subsection{Deriving transit parameters}

To derive the transit parameters of each simulated dataset, we used the best GP noise model, as derived
above, or the G1 noise model using the transit model as the mean function. We also compared the performance of the GP models with a non-GP model. We chose a second-order polynomial detrending,  which is commonly used in the literature. We extracted the region of the light curves with three times the transit duration and centred in the mid-transit times. Then for each transit we fitted a second-order polynomial to the out-of-transit data and used it to normalise the transit. These normalised transits were fitted simultaneously with an  MCMC procedure similar to that explained above, but without the Kernel term, and considering only a white noise component. It should be noted that this procedure changes the data previous to the fit so the comparison with the GP models is not straightforward.

When preforming the transit retrieval we kept the limb darkening
and the transit period fixed to the injected values and fitted the mid-transit time, planet-to-star
radius ratio, normalised separation of the planet, and its orbital inclination. The priors used are
given in Table~\ref{prior_transit}. We impose a prior on the inclination to insure that the impact
parameter is lower than one and prevent very grazing transits. Allowing grazing transits leads to a
high degeneracy between the parameters of the transit and prevents an efficient exploration of the
parameter space. Moreover, preventing very grazing transits also allows  the transit
detection to be estimated by analysing only the significance of the derived planet-to-star radius ratio. The priors
for the hyper-parameters of the GP noise model used were the same as in the previous section, but we
started the chains close to the best solution found previously to speed up convergence. We found
that if we started the chains of the hyper-parameters randomly, as before, the convergence could be
very slow, especially for the Earth-size planet, although it eventually reached the same result. 
When using real data, our procedure can be emulated by first fitting the GP noise model to the
out-of-transit data. Alternatively, very long chains are required.  For the non-GP model we used the same prior for the white noise component as  for  the GP model. The above MCMC procedure and selection of chains was used to estimate the parameters and uncertainties.

\begin{table} 
\caption{Priors for the fitted transit parameters. 
\label{prior_transit} }
\begin{tabular}{lcc} 
\hline 
\hline Parameter & Prior \\ 
\hline T$_{0}$ (days)   &  $\mathcal{U}(-0.1;0.1)$ \\ 
$R_p/R_{\star}$  & $\mathcal{U}( 0.000001;0.2)$ \\ 
$a/R_{\star}$  &  $\mathcal{J}(1.0; 30)$ \\ 
$inc$ [\degree]  &  $\mathcal{S}(i_\mathrm{graz};90)$  \\ 
\hline 
\hline 
\end{tabular} \\ 
$\mathcal{U}(a;b)$ is a uniform distribution between $a$ and $b$;  $\mathcal{J}(a;b)$ is a
Jeffreys distribution between $a$ and $b$; $\mathcal{S}(a,b)$ is a sine distribution between $a$ and
$b$; graz is the inclination corresponding to an impact parameter equal to 1 for a given 
$a/R_{\star}$. 
\end{table}

\subsection{Performance of the retrieval} 

\begin{figure*} 
\centering 
\includegraphics[width=0.90\columnwidth]{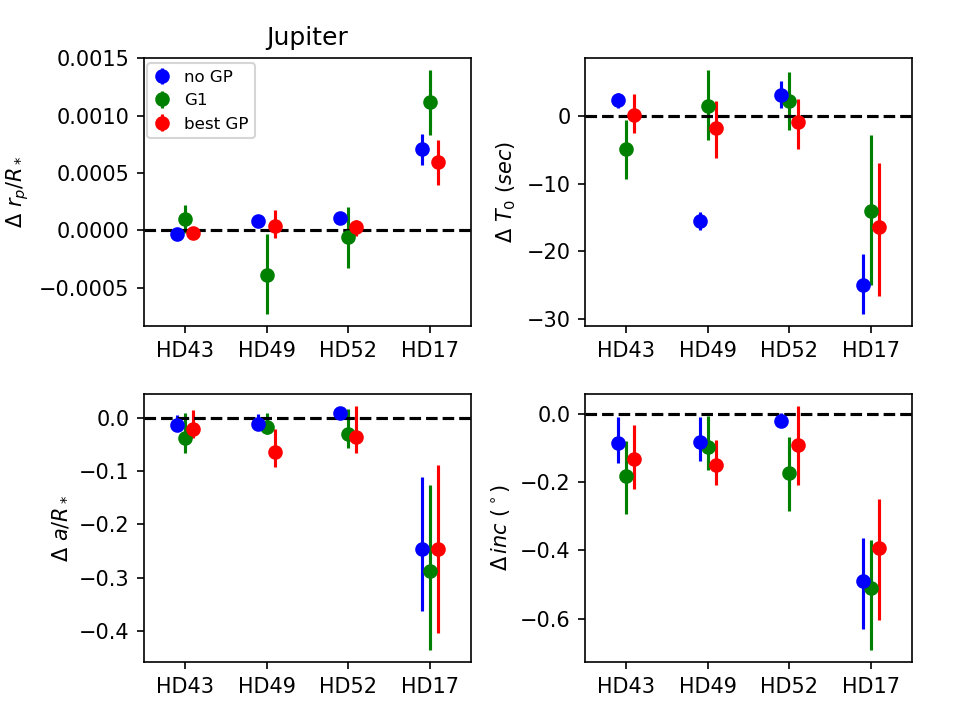}
\includegraphics[width=0.90\columnwidth]{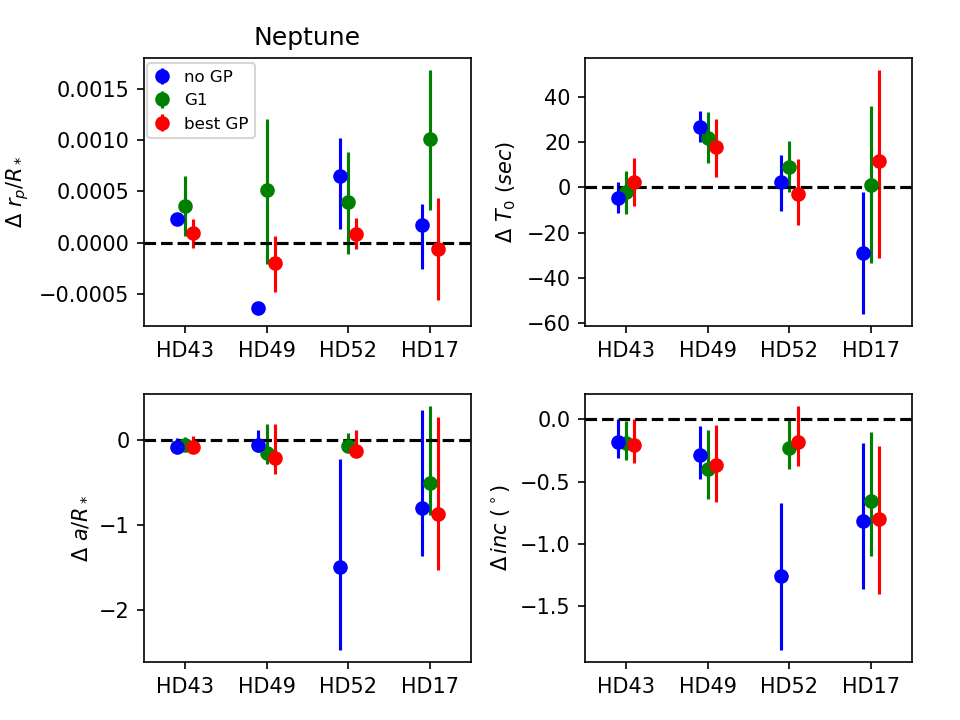} 
\caption{Difference between
the derived and  simulated parameters for all fitted parameters. Left: Results
for the Jupiter-size planet; Right:  Results for Neptune-size planet. Shown are the
G1 model (in green) and  the best GP noise model (in red). The true value is shown as a dashed line. \label{difplotbig}} 
\end{figure*}

\begin{figure} 
\centering 
\includegraphics[width=0.90\columnwidth]{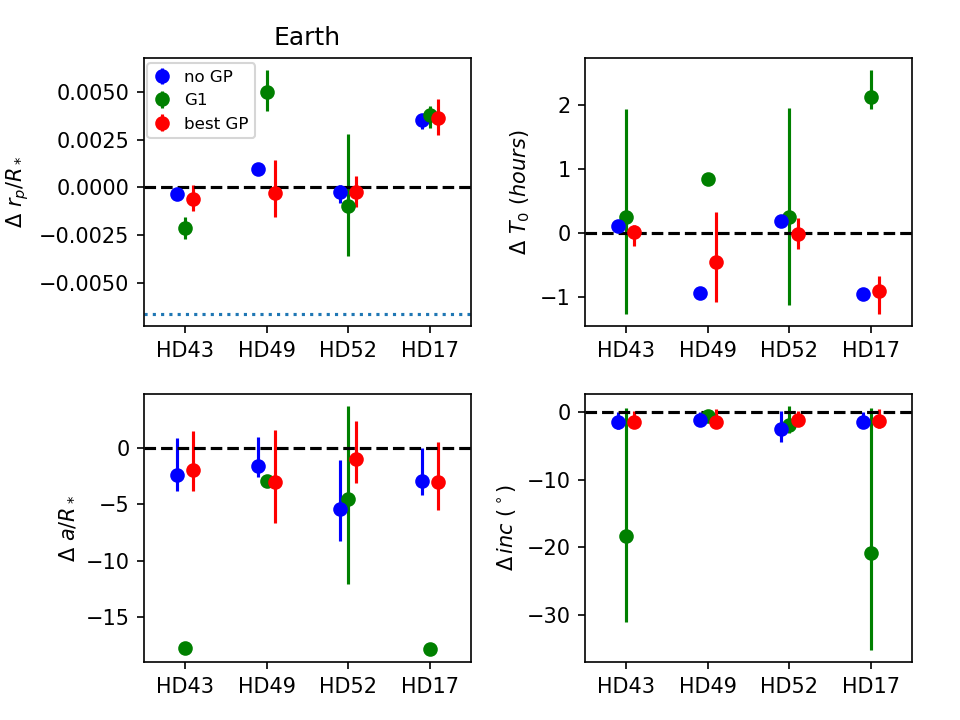}
\caption{Difference between the derived and the injected  parameters for the Earth-size planet.  Shown are the G1 model (in green) and  the best GP noise model (in red). The dotted line shows the value of the
injected planet-to-star radius ratio (implying  that a depth measurement 3 $\sigma$ away from this
line is consistent with a depth of zero and is therefore not significant). The unit of time is hours, and not seconds as for
the larger planets. The true value is shown as a dashed line. \label{difplot}} 
\end{figure}

For the best GP noise model, for the G1 model, and for the non-GP model, we
derived the difference between the estimated parameters and the injected ones. These differences are
shown in Figure~\ref{difplotbig} for the Jupiter- and Neptune-size planets, and in
Figure~\ref{difplot} for the Earth-size planets. Furthermore, in Figure~\ref{difplot} we also show
the value of the injected  planet-to-star radius ratio in order to   visually access significant
detections.

For the Jupiter-size planet we found that for the GP models all parameters are within $3 \sigma$ of the simulated
values. The only exception is the sub-giant HD~179079, where the planet-to-star radius ratio and the inclination are biased  for the G1 model by $3.9 \sigma$ and $3.6 \sigma$, respectively. For the best  GP noise model (G2R1O1) the planet-to-star radius ratio is still sightly biased ($2.9 \sigma$), but all the remaining parameters are well retrieved. The uncertainties are in general slightly larger for the G1 model than for the best GP noise model. In contrast, we found that for the non-GP  model the derived planet-to-star radius ratio is biased for all the stars, while the derived transit time is biased for HD~49933 and HD~179079. This is mostly due to the derived errors being much smaller than for the GP models.
Hence, the non-GP model leads to biased results in some cases and the best GP noise model leads to more precise and accurate results.

For the Neptune-size planet  all the derived parameters of the GP models are within $3 \sigma$ of the injected values.
The uncertainties of  $R_p/R_{\star}$ are larger for the G1 model than for the best GP noise model,
but in general for the other parameters the uncertainties are similar for both models.  In contrast, we found that for the non-GP model the derived planet-to-star radius ratio is biased for HD~43587 and HD~49933, and the derived transit time is biased for HD~49933. This is due to the underestimation of the errors in the non-GP model. Hence, the GP models are more accurate than the non-GP model.

For the Earth-size planet the estimated
values for the light curve of  HD~52265 are within $3 \sigma$ of the injected values for all models except the transit time derived with the non-GP model. However, the derived
value of the planet-to-star radius ratio is not significant for the G1 model (  $ r_{p}/R_{\star} =  0.00563^{+0.00379}_{-0.00261}$), while the planet is
well retrieved for the best GP noise model and the non-GP model. For the light curve of
HD~43587, the G1 model is biased for both $r_{p}/R_{\star}$ and $a/R_{\star}$. The large mismatch of $a/R_{\star}$
is actually an indication that the G1 model did not find the correct signal of the Earth-size planet and could be, in fact, modelling stellar variability. For the best GP
noise model all of the parameters are unbiased and the correct signal of the Earth-size planet was
retrieved.   For the non-GP model the transit is well retrieved and there is only a bias in the derived transit time. For the light curve of HD~49933 (LRa01), with the G1 model
all the parameters are biased (up to 122 $\sigma$) except the inclination suggesting that the transit was mistaken for stellar activity. Using the best GP noise model all of the parameters are unbiased and the
correct signal of the Earth-size planet was found.   For the non-GP model the transit is well retrieved, but  the planet-to-star radius ratio and the transit time are both biased due to an underestimation of the errors. Finally, for  the light
curve of the sub-giant HD~179079, the G1 model derived parameters are very biased except the inclination, which is unconstrained. This indicates that the G1 model is strongly biased by stellar variability and the transit signal of the Earth-size planet was not found. The best GP model (G2R1O1) retrieved the transit, but the planet-to-star radius ratio is biased by $4 \sigma$ and the mid-transit time is biased by $3.8 \sigma$.  For the non-GP model  the planet-to-star radius ratio and the transit time are both biased due to an underestimation of the errors, but the transit is well retrieved. In this case the $a/R_{\star}$ and $inc$ errors are similar to or larger than the best GP model, and they are not biased.

In summary, for the Earth-size planet the G1 model did not detect the planet for HD~52265 and did not find the correct transit signal for the planet for HD~43587, HD~49933, and HD~179079 due to stellar variability. In contrast, the best GP noise model correctly retrieved the Earth-size although for the sub-giant star the parameters are biased. The non-GP model also correctly retrieved the Earth-size although the planet-to-star radius ratio and the mid transit time are biased for some of the stars. Somewhat surprisingly, the non-GP model retrieves  the Earth-size planet better than the G1 model,  probably because  the G1 model provides a poor description of the activity and has more freedom. The method we used for the classical approach changes the data, which can lead to bias, but it also restricts the parameter space and helps constrain the transit. For example, cutting the light curves will not allow the transit solutions found by the G1 model for HD~43587 and HD~179079. Centring the cut light curves in the correct mid-transit time might also help the performance of the non-GP model. Finally, the second-degree polynomial normalisation using the correct mid-transit time also helps the model. For real cases where both the cutting of the light curve and the normalisation is done without the knowledge of the correct transit time will lead to further biases in the derived transit parameters. Therefore, we conclude that a better description of the stellar activity is necessary in order to characterise Earth-size planets in high signal-to-noise, high cadence light curves and to favour the best GP model.

\section{Discussion and conclusions}

In this work, we tested whether Gaussian processes allow us to improve the characterisation of transit
parameters in cases where stellar variability is the dominant noise source in a light curve.  To
achieve this goal, we used a sample of five high cadence ($32\,$sec), high signal-to-noise observations
of four stars taken by the CoRoT satellite. We started by determining which and how many stellar
variability components were present in each light curve. We tested models
with a maximum of five variability components with a combination of stellar oscillations, granulation, and
rotation. Using model comparison,   for our sample we found that  the best GP noise model requires at least
four to five variability components contrasting with the common practice of using only one variability component for transit retrieval.  The
difference of marginal likelihood between the best model and the one-component model is extremely large, and
hence multi-component models are highly favoured.

For light curves with a derived rotation period, we found that the best GP model was composed of one oscillation component, 
two granulation components, and one rotation component (G2R1O1). For HD~43587, for which we do not constrain the rotational period, we found that the best GP model includes one oscillation component and four granulation components. In this case we attribute the longer timescale granulation component (period =15.4 days) to stellar activity without a clear sinusoidal signal  \citep{Harvey1985}, and hence it is  better described by the granulation kernel.

We found that for the best GP noise model the derived timescales are in qualitative agreement with results from asterosismology. Therefore, our model provides results consistent with our astrophysical knowledge of the star. The advantage of GPs is that the models are defined and applied in the time domain and hence they can be combined easily
with a transit model. In this way, GPs can be used to model the stellar variability simultaneously
with transit modelling. The alternative is to use a two-step approach where first we filter the stellar variability and then we preform transit modelling. However, for low signal-to-noise transits, filtering the stellar variability can deform or even remove the transits. Therefore, models that couple the transit model with the stellar variability model are needed for planetary characterisation.

For transit analysis, it is  more relevant   if the number of components used to describe stellar
activity affects the derivation of transit parameters. Hence, in the second part of this work we tested whether the best GP model found also improves transit parameter estimation relative to the one-component model.  For completeness, we also compared the GP models with a non-GP model. As a non-GP model we chose a second-degree polynomial detrending, which is commonly used in the literature. We recommend avoiding  overinterpretation of the comparison between a GP model and a non-GP  model because details of the non-GP model  chosen can  influence   the results and the conclusions. Furthermore, non-GP approaches change the data, which can  bias the transit model and in our case can help constrain it.  We tested a general case of a planet with a 15-day orbital period
so that several transits were present in the light curves. We also tested three different planet
sizes, Jupiter, Neptune, and Earth, to probe the different signal-to-noise regimes.

For a Jupiter- and Neptune-size planet,   the classical second-degree polynomial detrending leads to a bias in the determination of the planet-to-star radius ratio and the mid-transit time. Accounting for activity with a one-component GP results in larger error bars that unbiases the parameter derivation. Interestingly, when more GP components are added to better describe stellar activity the uncertainty of the parameters slightly decreases and hence the more complex model is more precise while maintaining accuracy. However, for large planets the best GP model is  significantly better than the one-component GP model only for the case of the sub-giant HD~179079 
 because sub-giants have longer timescales of variability and higher amplitudes, which is also why to date it has only been necessary to account for stellar variability for sub-giants and giants \citep{Dawson2014, Barclay2015, Grunblatt2016}.
 Therefore, we conclude that as long as several transits are observed and the signal-to-noise ratio of the transit is high (for Jupiter- and Neptune-size planets) a simple GP model is sufficient to correct stellar variability and allows us to derive unbiased planetary parameters.

  For the Earth-size planet we found that the non-GP model performs better than the G1 model. This is probably due to the data being modified by the normalisation procedure and  because the G1 model is not a good description of the stellar activity. Furthermore, as low signal-to-noise transits can be mistaken for stellar activity and the non-GP model assumes no stellar activity the transit is better retrieved. However, in real data this could lead to false detections if the existence and time of transit is not known a priori. 
   We also found that when including more components in the GP model, the stellar activity is better characterised leading to 
 significantly better results than the one-component model for all the targets. The multi-component GP model allows us to correctly retrieve the transit model (while the one-component model fails), and it leads to more accurate results then the non-GP model.
 Hence, we conclude that the multi-component model is necessary for low signal-to-noise
 transits. Therefore, in the case of small planets, we recommend  using a multi-component GP model
 in the transit analysis. A better characterisation of stellar activity leads to a much better planetary characterisation.  
  These results will be relevant
 for the analysis of transit light curves from TESS, CHEOPS, and PLATO.

We tested here a case of a 15-day orbital period where several transits were obtained.
We expect that in other cases, for example very short periods (where the ingress--egress timescale is shorter and similar to the stellar variability timescales), very long periods (due to the low number of transits), or single transits, having a better model for stellar activity will also be important. In future work we plan to explore the advantages and disadvantages of this framework in the characterisation of transiting planets in some of these cases. In particular, small planets with very few available transits (1-2)  will be relevant for the search of Earth-like planets with PLATO.

\begin{acknowledgements}
SCCB acknowledges support from  Funda\c{c}\~ao para a Ci\^encia e a Tecnologia (FCT) through Investigador FCT contract IF/01312/2014/CP1215/CT0004. O.D.S.D. is supported in the form of work contract (DL 57/2016/CP1364/CT0004) funded by national funds through FCT. F.P.~acknowledges support from fellowship PD/BD/135227/2017 funded by FCT (Portugal) and POPH/FSE (EC)
This work was supported by FCT through national funds (PTDC/FIS-AST/28953/2017) and by FEDER - Fundo Europeu de Desenvolvimento Regional through COMPETE2020 - Programa Operacional Competitividade e Internacionaliza\c{c}\~ao (POCI-01-0145-FEDER-028953) and through national funds (PIDDAC) by the grant UID/FIS/04434/2019. This publication was written in the framework of the International Team on "Researching the Diversity of Planetary Systems" at ISSI (International Space Science Institute) in Bern. We acknowledge the financial support of ISSI and thank them for their hospitality.

\end{acknowledgements}

\bibliographystyle{aa} 

\bibliography{susana}

\clearpage

\begin{landscape}
\begin{table} 
\centering 
\caption{Derived G2R1O1 model hyper-parameters. The white noise is always between 3.5-5 ppm. \label{table:k3rotmodel}}
\centering 
\begin{tabular}{lccccccccc} 
\hline 
\hline 
Star &  \multicolumn{2}{c}{rotation kernel}   & \multicolumn{3}{c}{oscillation kernel}   & \multicolumn{2}{c}{first granulation kernel}   &   \multicolumn{2}{c}{second granulation kernel} \\ 
&  amplitude & rotation period & $a_{osc}$ &$ \tau_{osc}$  &q1 & $ a_{gran1}$ & $ \tau_{gran1}$ &$ a_{gran2}$ & $ \tau_{gran2}$  \\ 
& ppm &  days & ppm &min &  & ppm &min & ppm &min   \\ 
\hline
HD 179079       & 134$\pm$34    &       17.6$\pm$6.6    &       25.5$\pm$3.2    &       13.99$\pm$0.27  &       2.30$\pm $0.62   &       104.5$\pm$2.6   &       42.8$\pm$2.9    &       53.9$\pm$6.3    &       588$\pm$89  \\ 
HD 49933 (IRa01)        &       237$\pm$27      &       3.23$\pm$0.17   &       6.46$\pm$0.27   &       8.687$\pm$0.080 &       1.80 $\pm$0.10       &          48.5$\pm$1.4 &       24.8$\pm$1.2  & 22.1$\pm$4.4    &       71$\pm$12         \\
HD 49933 (LRa01) &      293.14$\pm$24.39        &3.35$\pm$0.10& 6.58$\pm$0.17   &       8.411$\pm$0.050 &       1.495$ \pm$0.053       &50.9$\pm$1.1   &       26.63$\pm$0.90  &32.50$\pm$2.00 &       97$\pm$13\\ 
HD 52265        &       170 $\pm$ 23    &       11.6$\pm$2.4    &       4.83$\pm$0.33   &       7.491$\pm$0.051 &       4.48$\pm$0.47   &       46.7$\pm$1.4    &       7.40$\pm$0.26   &       52.12$\pm$0.85  &       29.73$\pm$0.90\\
\hline
\hline 
\end{tabular} 
\end{table}

\begin{table} 
\centering 
\caption{Derived G5 model hyper-parameters. The white noise is $4.116 \pm 0.002$ ppm
\label{table:k3model}} 
\centering 
\begin{tabular}{lccc cc cc cc cc} 
\hline 
\hline 
Star & \multicolumn{3}{c}{oscillation kernel}   & \multicolumn{2}{c}{first granulation kernel}   &   \multicolumn{2}{c}{second granulation kernel}&   \multicolumn{2}{c}{third granulation kernel} &   \multicolumn{2}{c}{fourth granulation kernel} \\ 
  & $a_{osc}$ &$ \tau_{osc}$  &q1 & $ a_{gran1}$ & $ \tau_{gran1}$ &$ a_{gran2}$ & $ \tau_{gran2}$& $ a_{gran3}$ & $ \tau_{gran3}$ &$ a_{gran4}$ & $ \tau_{gran4}$\\
& ppm &  min &  &ppm &min   & ppm &min & ppm &hours & ppm & days     \\ 
\hline 
\hline 
HD 43587  &               5.66$\pm$0.20 &       7.634$\pm$0.037&3.19$\pm$0.15   &       53.83$\pm$0.76& 21.59$\pm$0.59  &27.2$\pm$1.4   &       87$\pm$11&        45.2$\pm$1.8     &               13.98 $\pm $0.88&          191$\pm$23   &       15.4  $\pm$1.4       \\ 
\hline 
\hline 
\end{tabular} 
\end{table}

\clearpage 
\end{landscape}

\end{document}